\documentclass[journal,twocolumn]{IEEEtran}

\usepackage[dvips]{epsfig}
\usepackage[dvips]{graphicx}
\usepackage{amsmath,amsfonts,bm,amssymb,psfrag,ifthen,color,subfigure,algpseudocode,setspace,cite,stfloats,booktabs,float,url,indentfirst,tabularx,multirow,arydshln}
\usepackage[ruled]{algorithm2e}
\usepackage{hyperref}

\begin{document}
\title{Joint Beamforming Design for RIS-enabled Integrated Positioning and Communication in Millimeter Wave Systems}
\author{Junchang Sun, Shuai Ma, and Shiyin Li
\thanks{Junchang Sun and Shiyin Li are with the School of Information and Control Engineering, China University of Mining and Technology, Xuzhou 221116, China (e-mail: sunjc@cumt.edu.cn, lishiyin@cumt.edu.cn).}
\thanks{Shuai Ma is with the Peng Cheng Laboratory, Shenzhen 518055, China (e-mail: mash01@pcl.ac.cn).}
}

\maketitle
\begin{abstract}
	Integrated positioning and communication (IPAC) and reconfigurable intelligent surface (RIS) are emerging as crucial technologies for the future of wireless networks. In this paper, we propose a RIS-enabled IPAC framework with the assistance of the millimeter wave. Initially, we derive explicit expressions of the time-of-arrival (ToA)-based Cram\'er-Rao bound (CRB) and positioning error bound (PEB), serving as performance metrics for positioning. Then, we optimize the integrated system by  designing both active beamforming at the base station (BS) and passive beamforming at the RIS. This joint optimization problem aims to minimize the transmit power while satisfying the communication rate and PEB constraints. Subsequently, we propose an efficient two-stage method to solve this non-convex problem, incorporating  exhaustive search and semidefinite relaxation (SDR) techniques. Finally, simulation results reveal the effectiveness of the proposed RIS-enabled IPAC system, capable of ensuring reliable data rates and high-precision positioning across diverse transmission scenarios through adjustments of critical system parameters.
\end{abstract}


\section{Introduction}\label{Introduction}
\subsection{Backgrounds and Motivations}
Integrated sensing and communication (ISAC) is regarded as an important and promising technology for future beyond 5G (B5G) and 6G wireless networks \cite{Liu2022JSAC, Wei2022MCOM}. ISAC systems are designed  to provide both  robust communication services and efficient sensing capabilities simultaneously. To achieve this, radar functionalities are often integrated into these systems. Examples include joint radar and communication (JRC) systems \cite{Feng2020JCC} and joint communication and radar (JCR) systems \cite{Kumari2021JSTSP}. With the growing adoption of millimeter-wave technology,  there is an opportunity to leverage its advantages for the design of such integrated systems.

In the context of a millimeter wave-enabled integrated system, particular emphasis is placed on enhancing the positioning capabilities within the sensing domain. Thanks to its high bandwidth and temporal resolution, the millimeter wave offers promising prospects for both communication and positioning tasks. Hence, an integrated positioning and communication (IPAC) system, harnessing the potential of millimeter waves, is formulated in our previous work \cite{Sun2023TCOMM}. However, the propagation characteristics of millimeter wave signals are notably influenced by environmental factors, such as obstructions and interruptions caused by obstacles.  Such obstructions  severely impact the performance of ISAC and/or IPAC systems, often rendering them ineffective. To tackle  this challenge, the concept of a reconfigurable intelligent surface (RIS) is introduced as an innovative strategy and solution.

RIS, also known as an intelligent reflecting surface (IRS), has been considered as a revolutionary technology for wireless networks. Comprising an array of passive components, the RIS presents a novel approach. It operates by adjusting the phase shifts of incident electromagnetic waves, thereby reconfiguring the wireless environment and establishing  virtual line-of-sight (VLoS) links for signal transmission \cite{Penglu2021TVT, Liu2022TWC}. RIS-aided system enables the improvement of both communication quality and positioning performance, particularly in challenging and harsh environments \cite{Li2022TCOMM}. As a result, the exploration  of RIS-based IPAC millimeter wave systems holds  immense value and significance in the realm of wireless communication and positioning.

\subsection{Related Works}
Indeed, the domain of ISAC and/or IPAC systems has attracted significant attention recently. A multitude of studies have delved into this area, exploring the synergies and possibilities it offers. For instance,  Liu {\emph{et al.}} \cite{Liu2022JSAC} and Zhang {\emph{et al.}} \cite{Zhang2022COMST} conducted surveys, providing valuable insights into the opportunities and challenges presented by these integrated systems. The ISAC/IPAC allows for the concurrent utilization of resources, such as frequency bands and hardware components, for both positioning and communication signals. This approach addresses the spectrum scarcity issue and contributes to cost reduction \cite{Liu2022COMST}. Ayyar {\emph{et al.}} \cite{Ayyar2019RADAR} proposed a communication-centric spectrum coexistence scheme to design a communication-radar system. This innovative approach enhances the system's versatility by harmonizing communication and radar functionalities. Kumari {\emph{et al.}} \cite{Kumari2021JSTSP} developed a joint signal waveform strategy  for adaptive radar communication, thereby optimizing system efficiency. Additionally, Liu {\emph{et al.}} \cite{Liu2020ACCESS} delved into the spectrum sharing quandary in radar transmitter design. These efforts collectively highlight the potential and advantages of integrated designs in the context of communication, sensing, and positioning.

The rise of intelligent transportation, smart travel, and unmanned aerial vehicles (UAVs) has significantly heightened interest in IPAC systems utilizing millimeter wave technology. This interest is complemented by the growing deployment of millimeter wave base stations (BSs). Notably, the IPAC framework in conjunction with millimeter waves has spurred several notable studies and developments in recent times. Han {\emph{et al.}} \cite{Han2022JCC} introduced a data-aided positioning system,  thoroughly  exploring the trade-offs of the integrated millimeter wave system. The work offers valuable insights into the trade-offs inherent in these systems. In \cite{Sun2022LWC}, the authors proposed a robust beamforming scheme for IPAC millimeter wave systems, while considering the impact of  positioning errors. Kwon {\emph{et al.}} \cite{Kwon2021JSTSP} embarked on the joint design of beamforming and power allocation techniques, aimed at optimizing both communication and positioning aspects in IPAC setups. Further, in \cite{Sun2023TCOMM}, the authors formulated an IPAC model  incorporating  the Ziv-Zakai bound (ZZB) and undertook a comprehensive analysis of the trade-offs between communication and positioning facets.

However, the millimeter wave signal is sensitive to the environments, particularly obstacles and blockages. This phenomenon can result in degraded system performance, hindering the realization of effective communication and accurate positioning. To address these challenges, as mentioned above, RIS has emerged as a promising solution.  RIS introduces a mechanism to establish VLoS links, which can substantially aid the propagation of millimeter wave signals.  RIS-enabled systems exhibit notable improvements in both communication quality and positioning accuracy.
In the realm of communication, works such as those by  \cite{Wu2019TWC, Wu2019ICASSP} delved into active and passive beamforming strategies, thereby assessing the efficacy of RISs in enhancing communication performance.  Huang {\emph{et al.}} \cite{Huang2019TWC} investigated the energy efficiency of wireless networks, considering the incorporation of RIS components. Similarly, RIS technology has proven beneficial for positioning applications. Abu-Shaban {\emph{et al.}} \cite{Abu-Shaban2021ICC} designed a low-complexity positioning estimator that leverages RIS assistance. Elzanaty  {\emph{et al.}} \cite{Elzanaty2021TSP} derived the Cram\'er-Rao bound (CRB) expression for RIS-aided positioning systems and evaluated the impact of the RIS phase configurations  on theoretical bounds. Additionally, Keykhosravi {\emph{et al.}} \cite{Keykhosravi2022JSTSP} proposed the use of RIS for estimating the locations of mobile user equipment (UE) by optimizing RIS phase profiles.

In contrast to the individual analysis of communication and positioning performance described above, several works  focused on the design of RIS-aided integrated systems. For instance, Yu {\emph{et al.}} \cite{Yu2022TSP} introduced a RIS-enabled integrated framework with a focus on designing a novel transmission protocol. This approach shows the potential  of RIS in enhancing integrated systems through innovative transmission strategies. Luo {\emph{et al.}} \cite{Luo2023TVT} delved into signal-to-noises (SNRs) of  radar detection and user communication with the involvement of RIS. The results demonstrated the benefits that RIS can bring to both communication and radar sensing aspects. Moreover, Hu {\emph{et al.}} \cite{Hu2021LWC}  addressed the challenge of uncertain UE locations by proposing a robust RIS-aided communication system, and showed that the quality of service (QoS) requirement can be satisfied with the assistance of a RIS. 
Luan {\emph{et al.}} \cite{Luan2022VTC} derived the squared position error bound (PEB) within a RIS-enhanced system. Their work formulated an optimization problem that minimizes PEB while ensuring achievable rate constraints, offering insights into the potential of RIS to improve positioning accuracy. While these works have made significant strides in understanding the benefits of RIS within integrated systems, it is evident that there remains an avenue to explore the joint beamforming designs within IPAC systems that incorporate RIS.

\subsection{Main Contributions}
Inspired by the discussions presented earlier, in this paper, we propose a RIS-enabled IPAC  framework and jointly optimize the active and passive beamforming to simultaneously satisfy both the rate and PEB requirements. The main contributions of this paper are summarized as follows:
\begin{itemize}
	\item We propose a RIS-enabled IPAC framework within  a multi-user multi-input single-output (MU-MISO)  millimeter wave system. Moreover, we derive explicit expressions of both the CRB and PEB for each UE based on the time-of-arrival (ToA) estimation. The PEB is then adopted as the key positioning metric for guiding the integrated design.
	\item To improve the performance of our proposed IPAC systems, we formulate an optimization problem by optimizing the active beamforming at the BS and passive beamforming at the RIS. The problem aims to minimize the total transmit power subject to the achievable rate and PEB constraints. Specifically, the optimization concerning the passive beamforming takes into account two cases: discrete RIS phase shifts and continuous RIS phase shifts.
	\item We propose an efficient two-stage method to address the formulated problem. In the first stage, we focus on optimizing the RIS phase shift. For discrete RIS phase shifts, we employ an exhaustive search method to explore the available options. For continuous RIS phase shifts, we leverage the semidefinite relaxation (SDR) technique.  In the second stage, based on the obtained RIS phase, we again apply the SDR method to optimize the transmit beamforming in the BS.
	\item We conduct extensive simulations to demonstrate the performance of the proposed RIS-enabled IPAC system. Specifically, we set two different transmission obstruction scenarios to evaluate the system capability. The results show that the integrated system is effective in both the achievable rate and theoretical positioning error lower bound.
\end{itemize}

{\em Notations:} Boldfaced lowercase and uppercase letters represent vectors and matrices, respectively. $\Re \left\{  \cdot  \right\}$ denotes real part. $\left[ {\bm{a}} \right]_i$ denotes the $i$th element in the vector ${\bm{a}}$ and $\left[ {\bm{A}} \right]_{i,j}$ denotes the $\left( i,j \right)$th element in the matrix ${\bm{A}}$. The ${\rm{rank}}\left(  \cdot  \right)$, ${\rm{tr}}\left\{  \cdot  \right\}$, ${\left|  \cdot  \right|}$, ${\left\|  \cdot  \right\|}$, ${\left(  \cdot  \right)^{T}}$, ${\left(  \cdot  \right)^{H}}$, and ${\left(  \cdot  \right)^{-1}}$ denote rank, trace, absolute value, 2-norm, transpose, complex transpose, and inverse operations, respectively. ${\bm{A}} \succeq {\bm{B}}$ means that matrix ${\bm{A}} - {\bm{B}}$ is positive semidefinite. ${\mathbb E}\left\{  \cdot  \right\}$ and $\Pr \left\{  \cdot  \right\}$ denote the expectation and the probability operator, respectively.  ${\bm{I}}$ is the identity matrix and ${\bm{1}}_{N}$ is a $N\times 1$ vector with all elements being ones. The key notations and acronyms are listed in Table. \ref{notations} and Table \ref{acronyms}, respectively.
\begin{table}[H]
	\centering
	\small
	\caption{Summary of Key Notations}
	\begin{tabular}{|l|l|}
		\hline
		Notations &  Description \\
		\hline
		${\bm{w}}_{n,k}$ & Beamformning vector in the BS\\
		\hline
		$n$, $k$ & $n$th subcarrier and $k$th UE\\
		\hline
		${\bm{\Phi}}$ & RIS reflection-coefficient matrix \\
		\hline
		${\bm{v}}$ & RIS phase shift vector \\ 
		\hline
		${\bm{h}}_{{\rm{d}},n,k}$ & Channel from the BS to the UE \\
		\hline
		${\bm{h}}_{{\rm{r}},n,k}$ & Channel from the RIS to the UE \\
		\hline
		${\bm{G}}_n$ & Channel from the BS to the RIS \\
		\hline
		${\chi}_k$ & Obstruction indicator \\
		\hline
		${\gamma}_k$, ${\cal{P}}_k$ & SINR and PEB \\
		\hline
		$R_k$ & Achievable data rate \\
		\hline
		$q$ & Quantization bits number \\
		\hline
		$\Delta_f$ & Subcarrier spacing \\
		\hline
		${r}_k$ & Data rate requirement \\
		\hline
		$\delta_k$ & PEB threshold \\
		\hline
	\end{tabular}\label{notations}
\end{table}

\begin{table}[H]
	\centering
	\small
	\caption{Summary of Main Acronyms}
	\begin{tabular}{|l|l|}
		\hline
		Acronyms &  Description \\
		\hline
		ISAC & Integrated sensing and communication\\
		\hline
		IPAC & Integrated positioning and communication\\
		\hline
		RIS & Reconfigurable intelligent surface \\
		\hline
		MU-MISO & Multi-user multi-input single-output \\
		\hline
		CRB & Cram\'er-Rao bound \\
		\hline
		PEB & Position error bound \\
		\hline
		SINR & Signal-to-noise-plus-interference ratio\\
		\hline
		CSI & Channel state information \\
		\hline
		FIM & Fisher information matrix \\
		\hline
		EFIM & Equivalent Fisher information matrix \\ 
		\hline
		SDR & Semidefinite relaxation \\
		\hline
	\end{tabular}\label{acronyms}
\end{table}

\bibliographystyle{IEEE-unsorted}

\bibliography{refsRIS}

\end{document}